\documentclass{article}
\usepackage{amsmath, amssymb, graphics, setspace}
\usepackage{graphicx}
\usepackage{latexsym,epsfig,epsf,rotate}
\usepackage{hyperref}
\usepackage{xcolor}
\usepackage{epsfig, epstopdf}
\usepackage{hyperref}  

\begin{document}

\begin{titlepage}

\begin{flushright}
{ \bf IFJ-PAN-IV-2022-12 \\
  September 2022
}
\end{flushright}

\vspace{0.1cm}
\begin{center}
  {\Large \bf Spin correlations in $\tau$-lepton pair production due to anomalous magnetic and electric dipole moments}
\end{center}

 \vspace{0.1cm} 
\begin{center}
{\bf  Sw.~Banerjee$^{a}$,  A.Yu.~Korchin$^{b,c,d}$ and Z.~Was$^{d}$}\\
{\em  $^a$University of Louisville, Louisville, Kentucky 40292, USA} \\
{\em  $^b$NSC Kharkiv Institute of Physics and Technology, 61108 Kharkiv, Ukraine} \\
{\em  $^c$V.N. Karazin Kharkiv National University, 61022 Kharkiv, Ukraine} \\
{\em $^d$Institute of Nuclear Physics Polish Academy of Sciences, PL-31342 Krakow, Poland}\\
\end{center}
\vspace{.1 cm}
\begin{center}
{\bf   ABSTRACT  }   
\end{center}

We present a simple algorithm for the calculation of event weights
embedding the effects of anomalous electric and magnetic dipole  moments
in simulation of $e^-e^+\to \tau^-\tau^+ (n\gamma)$ events,
and the subsequent decay of the $\tau$ leptons produced.
The impact of these weights on the spin-correlation matrix and
the total cross section is taken into account.
The algorithm is prepared to work {\it in situ} the  Monte Carlo {\tt KKMC} program
without the need for introducing any external change to the generator libraries.
As an example, 
$e^-e^+ \to \tau^-\tau^+ (n\gamma), \; 
\tau^- \to \rho^- \nu_\tau \to \pi^- \pi^0 \nu_\tau, \ 
\tau^+ \to \rho^+ \bar{\nu}_\tau \to \pi^+ \pi^0 \bar{\nu}_\tau$
events were simulated at a center-of-mass energy of 10.58 GeV.
The distributions of the acoplanarity angle between
the planes spanned by the $\pi^- \pi^0$  and the $\pi^+ \pi^0$ momenta
of, respectively, $\rho^-$ and $\rho^+$ decays and 
in the rest frame of the entirely visible $\rho^-\rho^+$ system 
are presented for different values of the coupling constants
incorporating anomalous electric and magnetic dipole  moments
in the $\tau^- \tau^+ \gamma$ vertex.

\vfill %
\vspace{0.1 cm}

\vspace*{1mm}
\bigskip

\end{titlepage}

\vspace{1cm}
\section{Introduction}

The recently improved searches for electric dipole moments in $\tau$-lepton pair production~\cite{Belle:2021ybo}
has brought renewed attention on this topic in recent papers~\cite{Chen:2018cxt, Bernreuther:2021elu, Bernreuther:2021uqm}.
Furthermore, the deviation in measurement of anomalous magnetic moment of the 
muon~\cite{Muong-2:2021ojo} from its theoretical predictions has brought renewed attention for their measurement 
in $\tau$-leptons~\cite{Crivellin:2021spu}, where the new physics contributions could be 
enhanced, as mentioned in Ref.~\cite{Eidelman:2007sb} and references therein.
Because of the short lifetime of the $\tau$-lepton~\cite{Belle:2013teo}, the design of observables is complicated,
and measurable signatures need to be combined from many $\tau$-decay channels. 
Not only are all the final state particles from decays of the $\tau$-lepton not observed due to the presence of neutrinos,
the kinematic constraints from energy-momentum conservation also need to be modified 
due to the presence of initial-state bremsstrahlung photons, some of which are often lost in the beam pipe.
The use of the $\tau$-decay vertex position~\cite{Desch:2003mw} may be of help,
but it is not straightforward technique, since the details of experimental arrangements
are critically important information, and experimental resolutions due to different detector responses
need to be taken into account using simulations.

A short list of earlier searches for anomalous electromagnetic moments in the $\tau$-pair production in $e^- e^+$ collisions
include Refs.~\cite{L3:1998gov,OPAL:1996dwj,OPAL:1998dsa,ARGUS:2000riz,DELPHI:2003nah},
and the role of the electron longitudinal polarization were noted in Refs.~\cite{Bernreuther:1993nd,
Ananthanarayan:1994yb, Ananthanarayan:1994af, Ananthanarayan:1993yr}.
More recently, much attention has been paid to the LHC observations~\cite{ATLAS:2022ryk,CMS:2022arf}
for obtaining information on anomalous electromagnetic moments of the $\tau$ lepton
in peripheral collisions and $\gamma \gamma$ production of $\tau$ pairs~\cite{Dyndal:2020yen, Buhler:2022knp}, 
by using the phenomenon of spin precession in bent crystals~\cite{Fomin:2018ybj, Fu:2019utm}, 
and in the $\gamma p$ collisions~\cite{Koksal:2017nmy}.

At present, for Belle II phenomenology, the {\tt KKMC} Monte Carlo program~\cite{Jadach:1999vf}
is widely used in generation of event kinematics as input to detailed simulation of detector response.
The program features all major $\tau$-decay channels, complete with spin effects including longitudinal
and transverse spin correlations between the two $\tau$-leptons, higher order QED
corrections, and capable of assuring that the $e^-e^+ \to \tau^- \tau^+ (n \gamma)$
cross-sections reach precisions at the per mille level~\cite{Banerjee:2007is}.
Effects of anomalous dipole moments are not expected to be large,
but the development of necessary calculations and tools which ensure the high precision
of the Standard Model (SM) prediction is at present of renewed interest.
Numerical effects of  anomalous couplings need to be presented
in the form in which all SM effects as well as detector resolution and acceptance effects
are taken into account at the same time with sufficient precision.
That is why it is useful to prepare theoretical predictions 
in a form suitable for use with Monte Carlo event generation programs.
Here, we present such a solution for the {\tt KKMC} program.

Our paper is organized as follows. In Section~\ref{sec:ampl}, we present the amplitudes and spin-correlation matrix.
The conventions and orientation of the quantization frames are emphasized.
Compatibility with respect to the choices used in
Refs.~\cite{Jadach:1984iy,Jadach:1984hwn,Przedzinski:2018ett,Jadach:1998wp}
are discussed in Section~\ref{sec:numer}, where some numerical results are discussed as well.
First group of results is oriented toward tests of the algorithm.
For these results, all final-state momenta including the unobservable $\tau$-neutrino momenta are used.
We next discuss the results with a semirealistic observable, which is the acoplanarity angle $\varphi$
between the two planes spanned by the decays $\rho^- \to \pi^- \pi^0$ and 
$\rho^+ \to \pi^+ \pi^0$, and defined in the  $\rho^- \rho^+$ rest frame.
The events of the process
$e^-e^+ \to \tau^-\tau^+ (n\gamma), \; 
\tau^- \to \rho^- \nu_\tau \to \pi^- \pi^0 \nu_\tau, \ 
\tau^+ \to \rho^+ \bar{\nu}_\tau \to \pi^+ \pi^0 \bar{\nu}_\tau$
are used to monitor the impact of anomalous magnetic and electric dipole moments.
In Section \ref{sec:program}, we demonstrate how the algorithm can be installed and 
used  for weights of  events simulated with the {\tt KKMC} program.
The summary Section~\ref{sec:summary} closes the paper.


\section{Amplitudes and spin-correlations}
\label{sec:ampl}

We consider electron-positron annihilation to a pair of $\tau$ leptons
\begin{equation}
e^- (k_-) + e^+ (k_+) \to \tau^- (p_-) + \tau^+ (p_+)
\label{eq:001}
\end{equation}
with the four-momenta satisfying energy-momentum conservation $k_- +k_+ = p_- + p_+ $. 

In the center-of-mass frame, the components of the momenta are  
\begin{eqnarray}
&& p_- =(E, \vec{p}), \qquad \;  p_+ =(E, \, -\vec{p}), \qquad  \vec{p} = (0, \, 0, \, p), 
\nonumber \\ 
&& k_- = (E, \, \vec{k}) , \qquad k_+  = (E, \, -\vec{k}),  \qquad 
\vec{k} =  (E \, \sin (\theta), \, 0, \, E \, \cos (\theta)), 
\label{eq:003}
\end{eqnarray}
so that the $\hat{z}$ axis is along the momentum $\vec{p}$, the reaction plane $\hat{x} \hat{z}$ is defined 
by the momenta $\vec{p}$ and $\vec{k}$, and the $\hat{y}$ axis is along $\vec{p} \times \vec{k}$. 
Here, $E$ is the beam energy related to the squared invariant energy $s$ by relation $E=\sqrt{s}/2$, 
$p = \sqrt{E^2-m^2}$ is the magnitude of the 3-momentum of the $\tau$ lepton,  
$m$ is the mass of the $\tau$, and $\theta$ is the scattering angle. In the following, we use the 
Lorentz factor $ \gamma=E/m$ and the $\tau$-lepton velocity $\beta = \sqrt{1-\gamma^{-2}}$.  
The mass of the electron is neglected.

The quantization frames of $\tau^-$ and  $\tau^+$ are connected to this reaction frame by 
the appropriate boosts along the $\hat z$ direction.  Note that the $\hat z$ axis  is parallel to 
momentum of $\tau^-$ but antiparallel to momentum of $\tau^+$. The momenta of the beams reside 
in the  $\hat x \, \hat z$ plane. Only the reaction frame, the $\tau^-$ and the $\tau^+$ rest 
frames are used for calculations throughout the paper. The vector indices used in the program 1, 2, 3, 4 
correspond to the $\hat x,~ \hat y,~ \hat z,~\hat t $ directions, respectively.

We assume that the electromagnetic vertex for the $\tau$ lepton has the following structure:  
\begin{equation}
\Gamma^\mu  = \gamma^\mu + \frac{\sigma^{\mu \nu} q_\nu}{2 m} \,  \bigl[ i  a(s) + \gamma_5 b(s) \bigr],
\label{eq:005}
\end{equation}
where $q = p_- +p_+$. 
The functions $a(s)$ and $b(s)$ are related to the Pauli and electric dipole form factors, respectively,
\begin{equation}
a(s) = F_2 (s),  \quad \qquad b(s) = F_3(s).
\label{eq:006}
\end{equation}
The form factors at $s \ge 4 m^2$ acquire imaginary parts due to loop corrections, and final-state 
interaction\footnote{In general, an imaginary part arises at $s>0$ due to the three-photon intermediate state 
in the higher-order loop corrections.}. 
Thus we choose the $a(s)$ and the $b(s)$ coefficients to be complex numbers. 
Up to two-loop corrections, one can use for the Dirac form factor $F_1(s)$, an approximation 
$F_1 (s) = 1$, and therefore,  we do not include $F_1(s)$ in the first term of Eq.~(\ref{eq:005}).

The quantity $a(0)$ is the anomalous magnetic dipole moment (AMDM), while $b(0)$ is related to the 
electric dipole moment (EDM) $d$, namely  
\begin{equation}
a(0) = \frac{1}{2} (g-2),    \qquad \quad b(0) = \frac{2m}{e} d.
\label{eq:007}
\end{equation}

In order to separate the contribution from New Physics (NP), one can explicitly include 
contribution from QED in $a(s)$. To the first order in the fine-structure constant $\alpha$,  
one has at $s \ge 4 m^2$  \cite{Berestetskii:1982}
\begin{equation}
a(s)_{QED}=\frac{\alpha m^2}{\pi \, s \, \beta} \, \bigl(  \log \frac{1-\beta}{1+\beta} + i \, \pi \bigr).
\label{eq:008}
\end{equation}
This contribution at very large values of $s $ ($s \gg 4 m^2$) behaves as 
\begin{equation}
a(s)_{QED} = \frac{\alpha m^2}{\pi \, s}\, \bigl( -\log \frac{s}{m^2} + i \, \pi \bigr).
\end{equation}

Therefore, in general we can write 
\begin{equation}
a(s) = a(s)_{QED} + a(s)_{NP}, \qquad \quad b(s) = b(s)_{NP}, 
\label{eq:009}
\end{equation}  
neglecting very small contribution to $b(s)$ in the SM.    

The matrix element for our reaction process can be written in terms of the spinors of initial- and 
final-state fermions as
\begin{equation}
{\cal M} = -\frac{e^2}{s} \bar{v}_e (k_+) \gamma_\mu u_e (k_-) \, 
\bar{u}_\tau (p_-) \Gamma^\mu v_\tau (p_+)
\label{eq:004}
\end{equation}
where $e$ is the positron charge satisfying $e^2  = 4 \pi \alpha$. 
The differential cross section is related to the matrix element squared through  
\begin{equation}
\frac{d \sigma}{d \Omega} = \frac{\beta}{64 \pi^2   s } |{\cal M}|^2.
\label{eq:011}
\end{equation}

We consider production of the polarized $\tau^-$ and $\tau^+$ leptons, which are characterized   
by the polarization 3-vectors in their rest frames, respectively 
\begin{equation}
\vec{s}^{\, -} = (s^-_1, \, s^-_2, \, s^-_3),  \qquad  \vec{s}^{\,+}=   (s^+_1, \, s^+_2, \, s^+_3),
\label{eq:012}
\end{equation}
where the Cartesian components are defined with respect to the chosen frame 
constructed from  $\vec{p_\pm}$, $\vec{k_\pm} $ momenta and their vector products $\vec{p_\pm} \times \vec{k_\pm}$.  
It is convenient to introduce the 4-th components of the spin vectors as follows: 
 \begin{equation}
s^-_i = (s^-_1, \, s^-_2, \, s^-_3, \, 1 ), \qquad s^+_j =   (s^+_1, \, s^+_2, \, s^+_3, \, 1 )
\label{eq:013}
\end{equation}
 with $i, ~j = 1,~2,~3,~4$.

After squaring the matrix element and averaging over the polarizations of electron and positron, 
we obtain an expression in the form:
\begin{equation}
|{\cal M}|^2 = \sum_{i, j=1}^4 \, R_{i j} \, s^-_i  s^+_j.
\label{eq:014}
\end{equation}
We keep only terms linear in $a(s)$ and $b(s)$.
 
The spin-correlation coefficients $R_{ij}$ listed below depend on the energy $s$ and the scattering angle $\theta$,
since $s$ and $\cos\theta$ differ for each event and its effective Born kinematic interpolation.
However, for brevity of notation, the energy and scattering angle dependence is not explicitly indicated,
and we denote $a(s)\equiv a$ and $b(s)\equiv b$ in the following expressions:
\begin{eqnarray}
&& R_{11} = \frac{e^4 }{4 \gamma ^2} \bigl(4  \gamma ^2 \, {\rm Re}(a) +\gamma ^2+1 \bigr) \sin ^2(\theta ), 
\nonumber \\
&& R_{12} = -R_{21} =   \frac{e^4}{2} \beta \sin ^2(\theta ) \, {\rm Re}(b),   \nonumber \\
&& R_{13} = R_{31} =  \frac{e^4 }{4 \gamma } \left[ ( \gamma ^2+1 ) {\rm Re}(a)+1\right] \sin (2 \theta ),
\nonumber \\ 
&& R_{22} = -\frac{e^4}{4 }  \beta^2 \sin ^2(\theta ), \nonumber \\
&& R_{23} = -R_{32} = - \frac{e^4}{4} \beta \, \gamma  \sin (2 \theta ) \, {\rm Re}(b),  \nonumber \\
&& R_{33} = \frac{e^4}{4 \gamma ^2} \left[ \bigl(4 \gamma ^2 \, {\rm Re}(a) +  \gamma^2 + 1\bigr) 
\cos ^2(\theta ) + \beta^2 \gamma ^2 \right], \nonumber \\
&& R_{14} = -R_{41} =  \frac{e^4}{4} \beta \, \gamma \sin (2 \theta )  \,  {\rm Im}(b),  \nonumber \\
&& R_{24} = R_{42} =  \frac{e^4 }{4 } \beta^2 \, \gamma   \sin (2 \theta ) \, {\rm Im}(a),  \nonumber \\
&& R_{34} = - R_{43} = - \frac{e^4}{2} \beta  \sin ^2(\theta ) \, {\rm Im}(b),   \nonumber \\
&& R_{44} = \frac{e^4 }{4 \gamma ^2} \left[4  \gamma ^2 \, {\rm Re}(a)  + \beta^2 \gamma ^2  \cos^2(\theta ) + \gamma ^2+1 \right].
\label{eq:015}
\end{eqnarray}

The imaginary parts of $a$ and $b$ lead to terms linear in   $\vec{s}^{\, -}$ and $\vec{s}^{\, +}$  which 
correspond to nonzero polarizations of the $\tau$ leptons. The polarizations along the $\hat x$
axis (called the transverse polarization) and  $\hat z$ axis (called the longitudinal polarization) 
induced by ${\rm Im}(b)$, are opposite in sign for $\tau^-$ and $\tau^+$, while those 
along the $\hat y$ axis (called the normal polarization) induced by ${\rm Im}(a)$ have the same sign. 

It is of interest to separate the contribution from the coupling $b(s)$, 
\begin{eqnarray}
|{\cal M}|^2_{\rm EDM} &=& \frac{e^4}{2} \beta  \,
 \{ \left[ (s^-_1 s^+_2 - s^-_2 s^+_1) \sin(\theta) -   \gamma  (s^-_2 s^+_3 - s^-_3 s^+_2) 
\cos(\theta) \right]  {\rm Re}(b)  \nonumber \\
&& - \left[ (s^-_3 - s^+_3) \sin(\theta) - \gamma (s^-_1 - s^+_1) \cos(\theta) \right] {\rm Im}(b) 
\} \sin (\theta),
\label{eq:15EDM}
\end{eqnarray}
which can be useful in determination of $b(s)$.  Recently new constraints on the value of EDM 
have been obtained by the Belle experiment at the KEKB $e^+ e^-$ collider~\cite{Belle:2021ybo}. 
These results improve upon the previous limits on EDM also obtained by Belle collaboration~\cite{Belle:2002nla}.  


Finally, we address the issue of radiative corrections.  
In {\tt KKMC} program, beyond exclusive exponentiation, QED corrections up to
the second order (real and virtual) are included or considered.
The two-loop virtual corrections, not enhanced logarithmically, are not
necessary in full, and are taken into account only in part.
In particular, the imaginary parts of the second-order contributions
to the form factors belong to that ignored category.
A more complete description of radiative corrections in {\tt KKMC}
and program limitations for use at relatively low energies is
given in Ref.~\cite{Banerjee:2007is}, 
where approximations for the box diagrams contributions, and
some other terms, also  proportional to $\alpha  m /\sqrt{s}$,
are discussed.  To reduce the ambiguities further, studies that are broader than those of
Ref.~\cite{Banerjee:2007is} would be needed. This remains
outside the scope of the present paper. 



\section{Numerical results and tests}
\label{sec:numer}

We first need to prepare  check if Eqs.~(\ref{eq:015}) are properly installed in our program
for the case $a(s)= b(s) = 0$. In this case, we obtain the expression
\begin{eqnarray}
|{\cal M}|^2_{a=b=0}  &= & \frac{e^4}{4 \gamma^2} \bigl\{  \gamma^2+1 + \beta^2 \gamma^2 \cos^2(\theta)
\nonumber \\
&& +s^-_3 s^+_3  \left[ \beta^2 \gamma^2      + (\gamma^2+1) \cos^2(\theta)  \right] \nonumber \\ 
&& + \left[s^-_1 s^+_1 (\gamma^2+1) -  s^-_2 s^+_2  \, \beta^2 \gamma^2  \right] \sin^2(\theta) \nonumber \\
&& +(s^-_1 s^+_3 + s^-_3 s^+_1) \, \gamma \sin(2 \theta)  \bigr\},
\label{eq:016}
\end{eqnarray} 
which is consistent with Ref.~\cite{Tsai:1971vv}, and under the condition $E \gg m$   
reduces to the known decomposition in Ref.~\cite{Alemany:1991ki}  
\begin{equation}
|{\cal M}|^2_{a=b=0, \, E \gg m}  =  \frac{e^4}{4 } 
\bigl[  (1 + \cos^2(\theta)) (1 + s^-_3 s^+_3) + \sin^2(\theta) (s^-_1 s^+_1 - s^-_2 s^+_2)  \big].
\label{eq:017}
\end{equation}

These well established formulas provide good intuition, but we have to check
if conventions of our code for calculation of weights implementing
effects of anomalous dipole moments, and in particular, the orientation of the reference frame,
match what is used in  {\tt KKMC} and {\tt TAUOLA},
as described in Refs.~\cite{Jadach:1984iy,Jadach:1984hwn,Przedzinski:2018ett,Jadach:1998wp}.

For our application, we calculate the following weights for events generated with {\tt KKMC+TAUOLA}:
\begin{eqnarray}
wt_{spin}^{SM} &=& R_{ij}^{SM}h_i^- h_j^+ \, / R_{tt}^{SM}, \label{eq:A}\\
wt_{spin} &=& R_{ij}h_i^- h_j^+  \, / R_{tt} \, /wt_{spin}^{SM},   \label{eq:B}\\
wt&=& R_{tt} \, / R_{tt}^{SM} , \label{eq:C}
  \end{eqnarray}
where $i,~j = 1,~2,~3,~4$ and subscript ${tt}$ stands for $i=j=4$.  
The so-called polarimetric vectors $ h^-_i$ and $ h^+_j$ in Eqs.~(\ref{eq:A})-(\ref{eq:C}) 
depend on kinematics of respectively $\tau^-$- and $\tau^+$-decay products~\cite{Jadach:1993hs}. 
   
The $R_{ij}$ elements are now calculated as functions of $s$ and $\cos(\theta)$
which need to be interpolated from the kinematics of generated events.
Interpolation may be non-trivial for the case where hard bremsstrahlung photons
are present, as clarified in the next section with technical details.
$R_{ij}^{SM}$ is obtained from $R_{ij}$ by setting anomalous couplings $a_{NP}, \, b_{NP}$, 
as well as the QED dipole moment, to zero.

We present two sets of tests in the subsections below.

\subsection{Technical tests with no AMDM and EDM form factors}

For the first set of tests, we check if spin correlations of Eq.~(\ref{eq:A})
match those present in {\tt KKMC}. For that purpose, we generate two samples of
events: one with the spin effects included and the other with spin effects not included.
For the latter case, spin effects are instead implemented with the event wights of Eq.~(\ref{eq:A}). 
 The following events were used:
$e^-e^+ \to \tau^-\tau^+ (n\gamma), \; 
\tau^- \to  \pi^-  \nu_\tau, \ 
\tau^+ \to \pi^+ \bar\nu_\tau$.
The distributions obtained for the two approaches need to coincide. In our technical test
we  use unobservable momenta of $\tau^\pm$, for separation of the sample into subsamples where
hard-, soft-, or no-bremsstrahlung events are present, and for the definition of
distributions as well. 

For the preparation of test  distributions we perform the following steps:
\begin{enumerate}
\item  define $\hat n_3^\pm$ versors as following directions of $\tau^\pm$ in the lab 
  frame,
\item  define $\hat n_2^\pm$ versors as following direction of
  vector products  of $\tau^\pm$ with $e^\pm$ beam momenta, also in the lab 
  frame,
\item   define $\hat n_1^\pm=\hat n_2^\pm \times \hat n_3^\pm$,
\item decompose the $\pi^\pm$ momenta denoted by $\vec{q}^{\, \pm}$, respectively,
 in the $\hat n^+$ and $\hat n^-$ frames,
\item the components of $\vec{q}^{\, \pm}$ obtained this way are used to monitor
 each of the $R_{ij}$ elements. For histograming, the vectors $\vec{q}^{\, \pm} $ are
used,  and events with $\cos (\theta)$ positive and negative are taken separately.

\end{enumerate}

We note that the frames $\hat n^+$ and $\hat n^-$ are obtained from each other 
by rotation around $\hat y$ axis on angle 180$^\circ$.  These frames 
are convenient for preparing histograms and are used below.   

\vskip 1mm
\textit{\bf Born and soft photon case}
\vskip 1mm

\noindent
The {\tt KKMC} event sample with spin correlation included and with spin correlations 
absent, but our $wt_{spin}^{SM}$ is used, are compared. To be sensitive to linear in $\cos (\theta)$ terms,
we request that the $\tau^-$ momentum lies in the forward hemisphere ($\cos(\theta) \ge 0$).
We first performed such test for events in which no bremsstrahlung photons are present,
and later for the ones with soft/collinear photons present as well. There were no distinguishable
differences for spin correlations taken from  {\tt KKMC} and from our weights,
both in the Born approximation and if the soft photons determined by the condition on the $\tau^- \tau^+$ 
invariant mass $m^2(\tau^-\tau^+) / s \, > \, 0.98$, were allowed.

\vskip 1mm
\textit{\bf Hard  photon case}
\vskip 1mm

\noindent 
Now we turn our attention to events in which photons are present  
and choose the $\tau^-  \tau^+ (n\gamma)$ events in which $m^2(\tau^-\tau^+)$ 
was  in the $0.2-0.98$ range of the collision  Mandelstam variable $s$.
Most of these events include hard photons collinear to the beams,
but we still check the impact of the dominant QED bremsstrahlung effects on our weighting method,
and confirm that events of high $p_T$ photons do not contribute sizable ambiguities.

\begin{figure}[!ht]
\centerline{\includegraphics[scale=1.0]{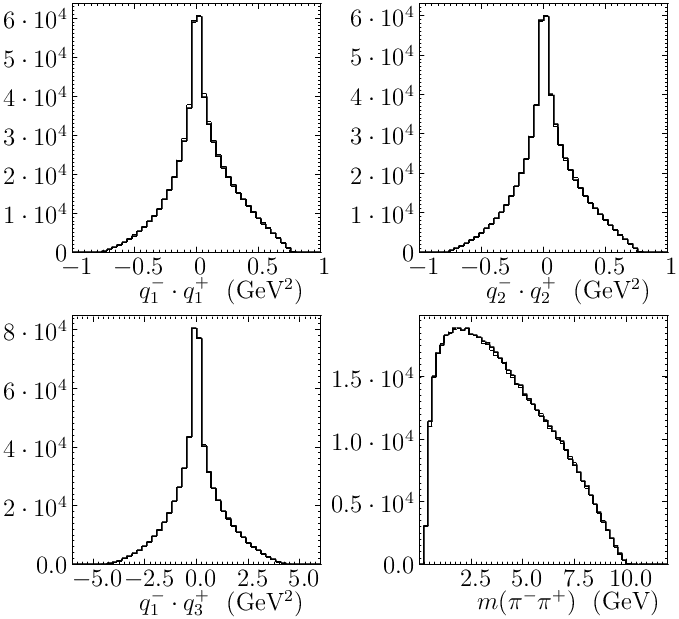}}
\centering
\caption{Correlation of $\pi^- \pi^+$ momenta (top left, top right and bottom left), 
and distribution over  $\pi^- \pi^+$ invariant mass (bottom right) for the $e^-e^+$ 
energy $\sqrt{s}=10.58$ GeV. 
Top panels: correlation of components  $x - x$ (left) and components $y - y$ (right); 
bottom: correlation of components $x - z$ (left).   
Anomalous couplings and QED contribution are not included. 
Two lines in each histogram show (i) {\tt KKMC} event sample with spin correlations included  
and (ii) {\tt KKMC} event sample without spin correlations, but instead $wt_{spin}^{SM}$ 
of Eq.~(\ref{eq:A}) is used.  The two lines of the histograms overlap almost completely.}
\label{fig:mpipi}
\end{figure}

The distribution of the invariant mass of the pion pair $m(\pi^- \pi^+)$ is shown in Fig.~\ref{fig:mpipi} 
(bottom right).  It is sensitive to the $z - z$ correlation, $R_{33}$. The $z - z$ correlation is 
strongly positive, this is because the $\tau$ leptons are relativistic. The  $z - z$ spin correlations 
are much better monitored by the distribution over $m(\pi^- \pi^+)$. 
Spin correlations enhance the number of events for $m(\pi^- \pi^+)$ around  $\sim$2~GeV
and less visibly around $\sim$8~GeV, as noted also in Fig.~2 of Ref.~\cite{Kaczmarska:2014eoa}.


The purposes of the plots presented are purely technical.
They are used to test elements of our code, first when photons are absent,
then for events where soft photons are allowed,
and finally, when only events of hard photons were selected.

For the case when $ m^2(\tau^-\tau^+)  / s >0.98$, where only soft photons may be present,
the comparison indicates statistically indistinguishable distribution.
Only the correlations $x-x$, $y-y$, $z-z$,  $x-z$ and $z-x$ are nonzero, as noted in Eq.~(\ref{eq:016}).
We also checked that other correlations are zero.

Thus, for the spin correlations of {\tt KKMC} and our spin weights,
only the plots with hard photons are included in the paper.
It should be noted that for the histograming purposes, the four momenta of only final states are used.

\subsection{Results with AMDM and EDM form factors}

Now we can turn our attention to distributions sensitive to the dipole moments.
Our aim is to check an extension to {\tt KKMC} event generator,
enabling inclusion of  AMDM and EDM form factors
through the event weights listed in Eqs.~(\ref{eq:B}) and (\ref{eq:C}).
Selection of the actual distribution is not straightforward.
For example, the $\nu_\tau$ and the $\bar{\nu}_\tau$ momenta are not observed directly, 
but partially can be inferred through the reconstruction of the $\tau$-decay vertex position.
This is, of course, $\tau$-decay channel dependent.
That is why the appropriate study including detection ambiguities
require simulation details of specific detector configurations,
and is out of the scope of this phenomenological paper.
Our emphasis here is devoted to extension of the {\tt KKMC} event generation tool,
a necessary prerequisite step to experimental studies involving detector simulations.

More than 25 years ago, it was demonstrated in Ref.~\cite{Kuhn:1995nn}
that the analyzing power of multi-meson final states in semileptonic $\tau$
decays with respect to the $\tau$ spin is equal and maximal for all decay modes.
Obviously the $\tau^\pm \to \pi^\pm \nu_\tau$ decay mode provides
signatures which are the easiest to interpret.
In this connection, we should mention the studies in Ref.~\cite{Bernabeu:2004ww, Bernabeu:2006wf, 
Bernabeu:2007rr, Bernabeu:2008ii} of the $\tau$-spin correlations in   
the $e^- e^+ \to \tau^- \tau^+ \to h^- \nu_\tau \, h^+ \bar{\nu}_\tau $ process,
where $h^\pm$ are hadrons and no secondary decays of $h^\pm$ were taken into account. 
The authors constructed asymmetries sensitive to AMDM and EDM   
for the conditions of Super B/flavor factories. Inclusion of the electron beam polarization 
was essential for isolating the $\gamma \tau^- \tau^+$ anomalous form factors.          

The main ambiguity of the measurement of the $\tau^\pm \to \pi^\pm \nu_\tau$ channel, and perhaps 
of  any two-particle decay channel $\tau^\pm \to h^\pm \nu_\tau$, may come from the precision 
of the $\tau$-decay vertex  position reconstruction.     
From this point of view the channel $\tau^\pm \to \pi^\pm \pi^\pm \pi^\mp \nu_\tau$ may be better.
On the other hand, systematic ambiguity from modeling of this $\tau$-decay channel may need to be revisited.
Thus the work of Ref.~\cite{Was:2015laa} may need to be revised for the new
application.

Our observable is independent of the details of the hadronic spectrum,
as long as the scalar contribution to hadronic current is negligible,
which was shown by the Belle experiment for $\tau$ decays to two scalars in Ref.~\cite{Belle:2008xpe}.

For these reasons, we have chosen to demonstrate the functionality of our code
with the $\tau^\pm \to \rho^\pm \nu_\tau \to \pi^\pm \pi^0  \nu_\tau$ channel.
We may rely, in preparation of our test observable, only on the kinematic of secondary $\rho$ decay.
Then there is no need to use $\tau$-decay vertex reconstruction, and that is why
this decay channel is of interest from the point of view of systematic ambiguity evaluation.

To demonstrate potential of our approach,
let us consider the cascade process:
$e^-e^+ \to \tau^-\tau^+ (n\gamma), \; 
\tau^- \to \rho^- \nu_\tau \to \pi^- \pi^0 \nu_\tau, \ 
\tau^+ \to \rho^+ \bar{\nu}_\tau \to \pi^+ \pi^0 \bar{\nu}_\tau$.
We use the method of Refs.~\cite{Bower:2002zx, Desch:2003rw}.
For constructing observables sensitive to the $CP$ parity of the Higgs boson,
it was suggested to measure the acoplanarity angle $\varphi$
between the planes spanned on the decays $\rho^- \to \pi^- \pi^0$ and 
$\rho^+ \to \pi^+ \pi^0$ and defined in the  $\rho^- \rho^+$ rest frame.
The momenta of all the pions are to be measured,
which would yield the $\rho^-$ and $\rho^+$ momenta,
which are then boosted to the $\rho^- \rho^+$ rest frame.

In order to be sensitive to the transverse spin correlations, additional cuts need to be applied.
Accordingly, we apply the following constraint on the sign of the product $y_1 y_2 > 0 $, where 
\begin{equation}
y_1  = \frac{E_{\pi^-}  - E_{\pi^0}}{E_{\pi^-}  + E_{\pi^0}}, \qquad 
y_2 = \frac{E_{\pi^+}  - E_{\pi^0}}{E_{\pi^+}  + E_{\pi^0}}.
\label{eq:y1_y2}
\end{equation}

Here $y_1$ and $y_2$ are measured from the decay products of  $\rho^-$ and $\rho^+$, respectively,
and the energies of the pions are taken in the rest frame of the $\rho^-\rho^+$ pair.
Further details are provided in the original papers~\cite{Bower:2002zx, Desch:2003rw}.


\begin{figure}[!ht]
\centerline{\includegraphics[scale=1.0]{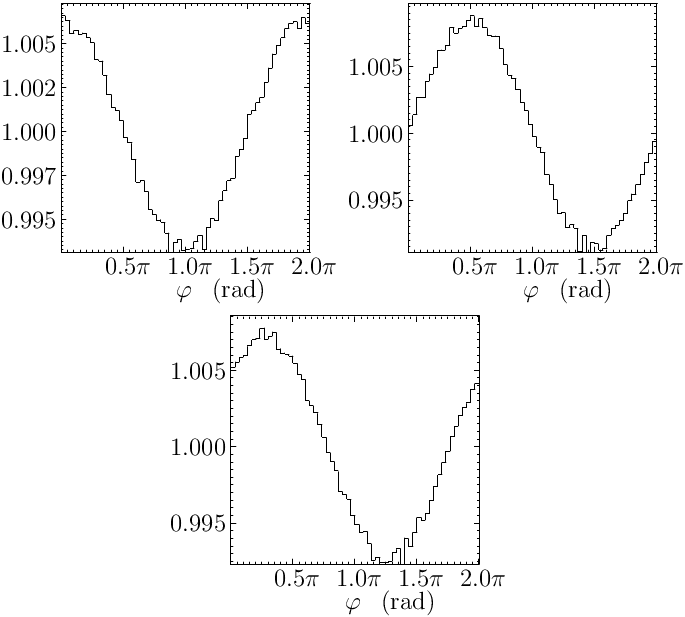}}
\centering
\caption{Distribution of $wt_{spin}$ as a function of the acoplanarity angle $\varphi$
at $\sqrt{s}=10.58$ GeV with the constraint $y_1 y_2 >0$. 
For the top left plot, ${\rm Re} (a_{NP}) = 0.04$ and other couplings are zero.
For the top right plot, ${\rm Re} (b_{NP}) = 0.04$ and other couplings are zero.
For the bottom plot, 
${\rm Re} (a_{NP}) = 0.04 \, \cos (\pi/4)$,  \  ${\rm Re} (b_{NP}) = 0.04 \, \sin (\pi/4)$ 
and other couplings are zero.}
\label{fig:ArBr}
\end{figure}
  	

In Fig.~{\ref{fig:ArBr}, we present results of the Monte Carlo simulation of the acoplanarity angle distribution using the weight method, for a few choices of the couplings $a(s)_{NP}$ and $b(s)_{NP}$.
We assume for simplicity that $a(s)_{NP}$ and $b(s)_{NP}$ take the real values.  
Of course, the absolute values are chosen arbitrarily and these figures demonstrate sensitivity of our method to the anomalous couplings. 

In these histograms, events without the photons or with the soft photons are selected using the constraint $m^2(\tau^- \tau^+) /s \ge 0.98$,
although our reweighting technique still works even if this criteria is relaxed.

It is seen from these figures, that the effect of the anomalous couplings on the distribution can reach 
about 0.005. Of course this is related to the absolute value of couplings taken 0.04,
which is already too large to be a realistic value.
This value is also much larger than the leading order QED correction 
$a(s)_{QED} = -0.000244+ i \, 0.000219$.

The distributions for ${\rm Re} (a_{NP})=0.04$ and ${\rm Re} (b_{NP}) =0.04$
have the form of a sinusoid and are shifted with respect to each other on the angle  $\sim \pi/2$.
The distribution for the case in which both couplings are present  
in Fig.~\ref{fig:ArBr} (bottom), is shifted on the angle $\sim \pi/4$ to the distributions in 
Fig.~\ref{fig:ArBr} (top).  It may be useful to combine ${\rm Re} (a)$ and ${\rm Re} (b )$ in one 
complex coupling
\begin{eqnarray}
&& c \equiv {\rm Re} (a) + i \, {\rm Re} (b) = 
\sqrt{({\rm Re}( a))^2 + ({\rm Re} (b ))^2} \, \exp{(i \psi)},  \label{eq:c(s)} 
 \\  && \tan (\psi) = {\rm Re} (b) / {\rm Re} (a). \nonumber
\end{eqnarray}           
Then it appears that the angle $\psi$ describes the shift of the distributions in Fig.~\ref{fig:ArBr}.
It gives a measure of $CP$ violation in the photon interaction with the $\tau$ leptons.  
Eq.~(\ref{eq:c(s)}) can be extended to the case in which $a $ and $b$ are complex. 


\section{Code implementation}
\label{sec:program}

One can, in future read in simulated events and use {\tt TauSpinner}
with pointer provided weights. One could also  
install $R_{ij}$ into Koralb, or one can play with {\tt Tralor} routine of KKMC
which define relation between $\tau$-lepton rest frames and laboratory frames.

At present, we assume that our solution will be applied  simultaneously
with {\tt KKMC} event generation and its internal variables will be available.
That is why we can proceed following technique presented in
Ref.~\cite{Jadach:1998wp}. The {\tt KKMC} polarimetric vectors $h^{\pm}_i$ are
first boosted to the laboratory frame using {\tt KKMC} routine specifically
prepared to the bremsstrahlung photons present in the event under consideration.
Also $h^{\pm}_i$ are available for all $\tau$-decay channels~\cite{Jadach:1993hs}. 
We simply need  to boost them back to the $\tau^\pm$ rest frames
but now of axes used in our code. Its relation with laboratory 
frame need to be controlled only.

Here we present with technical details related to {\tt basf2} version of {\tt KKMC},
as used in the Belle II software~\cite{Kuhr:2018lps}. 
In the file {\tt Taupair.fi} common block {\tt /c\_{}Taupair/} reside.
In this common spin polarimetric vectors for the first/second  $\tau$ can be found:
{\tt m\_{}HvecTau1(4) m\_{}HvecTau2(4)}.
They are needed for our weight calculation.
We copy these four vectors to user program variables and
boost them to lab frame using {\tt KKMC} {\tt  SUBROUTINE GPS\_{}TralorDoIt(id,pp,q)}
residing in {\tt GPS.F} file. These spin polarimetric vectors then need to be boosted
back to the rest frames of $\tau^-$ and $\tau^+$.
However now using routines from our application.
In this way the $\tau$-leptons rest-frames axes orientation is adjusted.

Now all informations our application need, and in its own reference frames, are available.

It is now straightforward to calculate $wt_{spin}$ and $wt$ of Eqs.~(\ref{eq:B}) and~(\ref{eq:C})
with $R_{i j}$ as a function of $s$ and $\cos(\theta)$, which are calculated from
the beam and $\tau$-lepton momenta in the $\tau$-pair rest frame,
with simple call to
$${\tt anomwt(iqed,Ar,Ai,Br,Bi,wtME,wtSPIN)}.$$
The {\tt iqed=1/0} activates/omits the QED SM part of anomalous magnetic moment.
The {\tt Ar, Ai, Br, Bi} denote respectively
real and imaginary parts of anomalous magnetic and electric dipole moments which are input for the 
calculation of {\tt wtME} and {\tt wtSPIN} weights representing
ratio of unpolarized cross section and spin effect factor for the anomalous
coupling included and absent. For each event such calculation can be repeated
several time for distinct numerical values of {\tt Ar, Ai, Br, Bi}.

This can be useful for fits and definition of optimal variables~\cite{Davier:1992nw,ALEPH:2013dgf}
for anomalous moment observables, or for Machine Learning applications.
We hope that the solution similar to the ones we attempted  in Ref.~\cite{Lasocha:2020ctd},
and references therein, can be used.

Note that routines from {\tt KKMC} are  used to obtain $h_i^-, \, h_j^+$ for each event.
Interpolation for $R_{ij}$ calculation  to the case in which the photons are present,
follows similar recipe as in~\cite{Richter-Was:2016mal}, where the
{\tt Mustraal} frame~\cite{Berends:1982ie} is used.


\section{Summary}
\label{sec:summary}

In this paper, we present a simple algorithm for the calculation of event weights
embedding effects of the anomalous magnetic and electric dipole moments in simulations
of $e^-e^+\to \tau^-\tau^+ (n\gamma)$ events with  the $\tau$ decays included.
Impacts on the spin effects and on the cross section are taken into account
with the help of two separate weights.

The algorithm is prepared to work with  {\tt KKMC} Monte Carlo program,  
without the need to introduce changes into generator libraries, but using
internal information from {\tt KKMC} common blocks. No internal information from {\tt KKMC}
is used, except through  calculation of $wt$ and $wt_{spin}$.
Solution is ready for use with  the Belle II software {\tt KKMC} installation~\cite{Kuhr:2018lps}.

The reweighting tool does not affect distributions measured from final-state
observable measured in real data from the experiment.
However, for example, by template fitting method, the experimental data can
be used to measure the fraction of dipole moment predicted by varying
the strength of the coupling related to dipole moments in {\tt KKMC} simulated samples.
This can confirm or rule out the strength of dipole moments
predicted by new physics models~\cite{Bernreuther:1996dr,Huang:1996jr}.

The option of reweighting previously generated events stored in datafiles
is possible, but as the dependence on additional couplings  is linear,
this may be  not necessary.
Solution like the one used in Ref.~\cite{Jacholkowska:1999ei} may be more convenient.
Our code  can be used to simultaneously calculate weights for four values of the dipole moments.
Thereafter, the weight for any other configuration can be obtained from their linear combination.

To demonstrate functioning of the algorithm, we calculate the effect of the
anomalous dipole moments on acoplanarity of two planes build with
$\pi^+ \pi^0$  and $\pi^- \pi^0$ momenta in the
$\pi^+ \pi^0 \pi^- \pi^0$ system rest frame of decay products 
in $e^-e^+ \to \tau^-\tau^+ n\gamma; \, \tau^\pm \to \pi^\pm \pi^0 \nu_\tau$
events. The numerical results are included, but we expect that, as in case of
the Higgs boson CP signatures, in this case also, the Machine Learning techniques
improving sensitivity and enabling combination of contributions
from all $\tau$-decay channels, can be efficient.  

Finally, we like to mention that the present method can be extended to the 
case of a polarized electron. This is important in view of the planned upgrade of 
the SuperKEKB $e^- e^+$ collider with polarized electron beam~\cite{Banerjee:2022kfy}.

\vspace{0.4cm}

\centerline {\bf Acknowledgments}

\vspace{0.4cm}

A.Yu.K. acknowledges partial support from the National Academy of Sciences of Ukraine 
via the programs  ``Participation in the international projects in high-energy and 
nuclear physics'' (Project No.~C-4/53-2022) and  ``Support for the development of priority 
areas of scientific research'' (6541230). 
This project was supported in part from funds of the Narodowe Centrum Nauki 
under Decision No.~ DEC-2017/27/B/ST2/01391.
This project was supported by the U.S. Department of Energy under Research Grant No.~DE-SC0022350.


\end{document}